\documentstyle[aps,prb,epsfig,twocolumn]{revtex}

\begin{document}

\bibliographystyle{prsty}

\draft

\wideabs{

\title{ Mathematical Structure of Evolutionary Theory }
\author{ P. Ao }
\address{ Departments of Mechanical Engineering and Physics, 
          University of Washington, Seattle, WA 98195, USA }
\date{ March 1, 2004 }

\maketitle

\begin{abstract}
   Here we postulate three laws which form a mathematical framework for the evolutionary dynamics in biology. The second law is most quantitative and is explicitly expressed in a unique form of stochastic differential equation. Salient features of Darwinian evolutionary dynamics are captured by this law: the probabilistic nature of evolution, ascendancy, and the adaptive landscape. Four dynamical elements are involved in the present formulation: the ascendant matrix, the transverse matrix, the Wright fitness function, and the stochastic drive. The first law may be regarded as a special case of the second law. It gives the reference point to discuss the evolutionary dynamics. The third law describes the relationship between the focused level of description to its lower and higher ones, and also defines the dichotomy of deterministic and stochastic drives.  
 A precise description of Wright's adaptive landscape is given and a new interpretation of Fisher's fundamental theorem of natural selection is provided. The generality of the proposed laws is supported by the demonstration of their equivalence to a general non-equilibrium dynamics formulation, based on a recently proved theorem. 
 Though the proposed laws are not the rigorous description, they should be regarded as mathematically what the laws of evolution might be. In addition, they provide a coherence framework to discuss several current evolutionary problems, such as speciation and stability, readily showing that statistical physics tools can be applied to Darwinian dynamics.     
\end{abstract}


}


\section{ Introduction }

Progresses in experimental biology and new data from field observation after the neo-Darwinian synthesis pose new questions to be answered and call the attention to previously unanswered questions. Biologists have been responding to this demand with tremendous activities, ranging from a critical discussing \cite{grene} of existing theories and courageous exploring \cite{kauffman} additional one, to the serious consideration of the epistasis of gene interactions \cite{epistasis}, to a reevaluation of shifting balance process \cite{coyne,goodnight,peck}, to a reexamining the concept of species \cite{pigliucci} and of the fundamental theorem of natural selection \cite{crow,grafen}, to various elegant mathematical models of speciation \cite{stewart,gavrilets,kisdi}. Even new philosophical implications were speculated \cite{es1998}. Such efforts have always enriched the theoretical and conceptual understanding of evolution. 

Among the various approaches, it has been noted \cite{epistasis} that a quantitative formulation of evolution dynamics may be of more importance to answer the new questions. Verbal description has been found to be inadequate, because many contributions to evolutionary dynamics, either independent or interactive, are of same magnitude, and the interactions can be highly nonlinear. It was from this consideration Stewart \cite{stewart} built his model based on symmetry-breaking and Gavrilets \cite{gavrilets} advanced his holey adaptive landscape model. This line of reasoning is further developed in the present article. 
Here we make an attempt to formulate a general and quantitative mathematical framework which appears broad enough to incorporate the ideas of Stewart \cite{stewart} and Gavrilets \cite{gavrilets} and is conceptually consistent with the Darwinian dynamics.  Two of the most influential concepts in evolutionary biology, the adaptive landscape and the fundamental theoretical of natural selection, are built naturally into the present formulation. The present work may also be regarded as an attempt to unify approaches from both biological and physical sciences. 

The present mathematical approach is based on the continuous approximation which treats the populations as continuous variable. 
This approximation has been well studied in biology, documented, for example, in both a commentated collection of historical articles \cite{li}, in a recent monograph \cite{burger} and in a recent textbook \cite{ewens}, and in an online book \cite{felsenstein}, and has been successfully employed in population genetics. In present article we will not elaborate further on this approximation. This implies that the equations to be discussed are of differential equation type. To be more precise, we will postulate three laws for evolution and the most important law, the second law, will be expressed in a unique form of stochastic differential equation. The connection of the present triad laws to previous results will be discussed.

We organize the rest of article as follows. In section II we postulate and discuss the three laws of evolution. Four dynamical elements are needed in this formulation. In section III the connection of the postulated three laws to previous formulations is discussed. An explicit computation of this connection is demonstrated in section IV. We also discuss the compatibility of present formulation with two current speciation models. The implications of present formulation were discussed in section V, and we conclude in section VI. 

\section{ Laws of Evolution }

In this section we postulate and discuss three laws on evolutionary dynamics. They form a quantitative mathematical framework for evolutionary dynamics. Then we discuss Fisher's fundamental theorem of natural selection and conclude that it is indeed an indispensable relationship. We start with the most important and quantitative law, the second law.

\subsection{ Second Law }

The central question naturally arises that how do we describe the evolutionary dynamics quantitatively and what are the dynamical elements? To be specific, let us consider an $n$ component biological system. The $n$ components may be the species in a evolutionary game \cite{maynardsmith}, or the traits to describe the speciation \cite{stewart}, or genes in the description \cite{kisdi}, or any quantities required to specify the system. The value of $j^{th}$ component is denoted by $q_j$. The $n$ dimensional vector ${\bf q}^{\tau} = (q_1, q_2, ... , q_n)$ is the state variable of the system. Here the superscript $\tau$ denotes the transpose. The dynamics of state variable is described by its speed $ \dot{\bf q}_t \equiv 
d {\bf q}_t /d t $ moving in the state space.

We postulate that the dynamics of the system is governed by following special form of stochastic differential equation, which consists of four dynamical elements, the semi-positive definite symmetric ascendant matrix $A$, the anti-symmetric transverse matrix $T$, the scalar function called Wright fitness function $\psi$, and the stochastic drive ${\bf \xi}$:
\begin{equation}
  [ A({\bf q}_t, t) + T ({\bf q}_t, t) ] \dot{\bf q}_t 
     = \nabla \psi ({\bf q}_t, t) + {\bf \xi}({\bf q}_t, t) \; ,
\end{equation}
and supplemented by the following relationship:
\begin{equation}
  \langle {\bf \xi}({\bf q}_t, t) 
          {\bf \xi}^{\tau} ({\bf q}_{t'}, t') \rangle 
    = 2 A({\bf q}_t, t) \; \epsilon  \; \delta(t-t') \; ,
\end{equation}
and $\langle {\bf \xi}({\bf q}_t, t) \rangle = 0$.
The connection of these two equations to conventional approaches will be discussed in next section. 
To ensure the independence of the dynamics of each component, we assume $\det [A({\bf q}, t) + T({\bf q}, t)]\neq 0$.
Here the subscript $t$ denotes that the state variable is a function of time, and $\delta(t)$ is the Dirac delta function.
In Eq.(2) we have assumed the stochastic drive is Gaussian white noise with zero mean. Factor 2 is a convention choice for the present formulation, and $\epsilon$ is a positive numerical constant, which for many situations might be set to be unity, $\epsilon = 1$, without affecting the biological description. The relationship between the stochastic drive and the ascendant matrix expressed by Eq.(2) guarantees that the ascendant $ A({\bf q}_t, t)$ is semi-positive definite and symmetric.
The average $ \langle ... \rangle $ is carried over the dynamics of the stochastic drive, and $\nabla$ is the gradient operator in the state space.

It is straightforward to verify that the symmetric matrix term is `ascendant': 
$ \dot{ \bf q}^{\tau}_t A({\bf q}_t, t) \dot{\bf q}_t \geq 0 $.
Its dynamical effect is to increase the fitness in terms of the Wright fitness function $\psi({\bf q}_t, t)$. Here we point out that the Wright fitness function defines the adaptive landscape. Hence ascendant matrix enables the systems to have the tendency to seek the largest fitness. This feature will be explicitly manifested in the discussion after the first law. 
The anti-symmetric matrix permits 'no change': 
$ \dot{\bf q}^{\tau}_t T({\bf q}_t, t) \dot{\bf q}_t = 0 $, therefore conservative. Dynamically it will not change the fitness.
A manifestation of the transverse dynamics is the oscillatory behavior.   
The dynamical effect of the stochastic drive ${\bf \xi}({\bf q}_t, t)$
on fitness is random: It may either increase or decrease the fitness.
With above interpretation, the static effect natural selection is represented by the gradient of fitness function, $\nabla \psi({\bf q}_t, t)$. The clear and graphical discussion of such Wright fitness function was first clearly expressed by Wright \cite{wright} in his concept of adaptive landscape, with which the present fitness function is identified. The tempo of natural selection is represented by the ascendant and transverse matrices. Eq.(1) states that the four dynamical elements, the gradient of fitness, the stochastic drive, the ascendant dynamics, and the transverse dynamics, must be balanced to generate the evolution dynamics. 

Eq.(1) is the fundamental equation of evolutionary dynamics expressed in a unique form of stochastic differential equation. 
In accordance with above discussion on stochastic drive and ascendant matrix, we may call the supplementary equation, Eq.(2), stochasticity-ascendancy relation, and will discuss it in the last subsection in this section in relation to Fisher's fundamental theorem of natural selection. 

The Wright fitness function $\psi$ is similar to a potential: It is in fact opposite in sign to the typical potential energy used in physical sciences. It is in general a nonlinear function of state variable. If it is further independent of time and is bounded above, the stationary distribution function $\rho({\bf q}, t=\infty)$ for the state variable, the probability density to find the system at ${\bf q}$ in state space, is expected to be a Boltzmann-Gibbs distribution:
\begin{equation}
   \rho({\bf q}, t=\infty) = \frac{1}{Z} \exp \left\{ 
                    \frac{ \psi({\bf q}) }{ \epsilon } \right\} \; ,
\end{equation}
with ${Z} = \int \prod_{i=1}^{n} d{q_i} \exp \left\{ {\psi({\bf q}) }
/{ \epsilon } \right\} $ the partition function, the integration over whole state space serving as the normalization factor.
Its justification will be given in next section. It is interesting to note that the dynamical aspects of evolution, the transverse and the ascendant matrices, do not explicitly show up in Eq.(3). From this expression we observe that the larger the constant $\epsilon$ is, the wider the equilibrium distribution would be, and more variation would be, or, the smaller the $\epsilon$ is, the narrower the distribution. In this sense we may call $\epsilon$ the evolution hotness constant.
The existence of such a Boltzmann-Gibbs type distribution suggests a global optimization. 

There are a few immediate and interesting conclusions to be drawn here. Near a fitness peak, say at ${\bf q}={\bf q}_{peak}$, we may expand the Wright fitness function, $\psi({\bf q}) = \psi({\bf q}_{peak}) - ({\bf q} - {\bf q}_{peak})^{\tau} U ({\bf q} -{\bf q}_{peak})/2 
+ O(|{\bf q} -{\bf q}_{peak}|^3) $. 
Here $U$ is a positive definite symmetric matrix as a consequence at the fitness peak. The stationary probability density to find the system near this peak is of a typical Gaussian distribution:  
\begin{equation}
  \rho({\bf q}, t=\infty) \propto \;  \exp \left\{ - \frac{
   ({\bf q} - {\bf q}_{peak})^{\tau} U ({\bf q} - {\bf q}_{peak}) }
   {2\epsilon} \right\} \; .
\end{equation}
Thus, away from the fitness peak, the probability to find the system will be exponentially small. Such a behavior has long been observed in many biological models \cite{burger,felsenstein}.

One may then wonder about how does the system move from one fitness peak to another? This process was first visualized by Wright \cite{wright}. The relevant mathematical calculation seems to be first done by Kramers \cite{kramers}. It had been applied to biology \cite{barton}, where it was shown that the stochastic drive must be involved. Quantitatively, the hopping from one peak to another must be aided by the stochastic drive. The dominant factor in the hopping rate $\Gamma$ is the difference in fitness between the peak and the highest point (saddle point ${\bf q}_{saddle}$) to cross the valley to another peak \cite{kramers,barton,vankampen}: 
\begin{equation}
 \Gamma \propto \exp \left\{ -  
   \frac{\psi({\bf q}_{peak}) - \psi({\bf q}_{saddle}) }
        {\epsilon }\right\}\; .
\end{equation}
This rate can easily be exponentially small. It is a quantitative measure of robustness and stability. Hence it may explain the usual observation, for example, that species is rather stable if viewing the peak as a definition for species. Nevertheless, Eq.(5) grants the possibility to hop between peaks when the stochastic drive is finite.  

The second law as expressed by Eq.(1) and (2) apparently capture the major features of the evolution dynamics first described by Darwin and Wallace \cite{darwin1858}, exhaustively exposed by Darwin \cite{darwin1958}, and further developed by Fisher, Haldane, Muller, and Wright \cite{founders}, and by many others \cite{burger,felsenstein}. Obviously it expresses the evolutionary process as a tinkering process \cite{jacob}. The necessity and chance \cite{monod} are represented by the Wright fitness function and the stochastic drive, respectively.

\subsection{ First Law }

The first law is a statement for the situation that there is no stochastic drive in the evolution, i.e., $\epsilon = 0$, the smallest possible value of the evolution hotness constant. This is clearly an approximation, because the variation is always there \cite{kimura}, though it may be regarded to be small and slow under certain time scale. Allowing the stochastic drive be negligible, Eq.(1) becomes
\begin{equation}
  [ A({\bf q}_t, t) + T ({\bf q}_t, t) ] \dot{\bf q}_t 
     = \nabla \psi ({\bf q}_t, t)  \; .
\end{equation}
Because of the ascendant matrix $A$ is non-negative, the system will approach the nearby attractor determined by its initial condition, and stay there forever. Specifically, because $ \dot{ \bf q}^{\tau}_t A({\bf q}_t, t) \dot{\bf q}_t \geq 0 $ and $ \dot{ \bf q}^{\tau}_t T({\bf q}_t, t) \dot{\bf q}_t = 0 $, Eq.(6) leads to 
\begin{equation}
  \dot{ \bf q}_t \cdot \nabla \psi ({\bf q}_t, t) \geq 0 \; .
\end{equation}
This equation implies that the deterministic dynamics cannot decrease the fitness: The speed of state variable $\dot{ \bf q}_t$ is in the same direction of the gradient of the Wright fitness function $\nabla \psi ({\bf q}_t, t)$. If the ascendant matrix is positive definite, i.e. $\dot{ \bf q}^{\tau}_t A({\bf q}_t, t) \dot{\bf q}_t > 0 $ for any nonzero $\dot{ \bf q}_t$, the fitness of the system always increases.
Hence, the first law clearly states that the system has the ability to find the local fitness or adaptive peak determined by the initial condition.

We have two remarks here. First, from the mathematical theory of dynamical systems, there are in general three types of attractors \cite{guckenheimer}: point, periodic, and chaotic (strange). The point attractors have been well explored in evolutionary biology since the work of Wright, corresponding the fitness peaks. Other two types of attractors have also been observed in biology \cite{murray,may}. Second, if we further assume the ascendant matrix $A=0$, the evolutionary dynamics will not change the system's fitness, hence is completely conserved. This is precisely what can be obtained from the Newtonian dynamics \cite{goldstein}. Based on this consideration one may incline to conclude that Newtonian dynamics is a special case of the Darwinian dynamics as expressed by Eq.(1) and (2), a conclusion many biologists may consider obvious \cite{dobzhansky,mayr}.  

The tendency implied in Eq.(6) to approach an attractor and to remain there has been amply discussed by Aristotle, well known in physics and biology. It is evident that in the present setting the first law is mathematically a special case of the second law. Nevertheless, this law does give us a reference point to define species and other relevant quantities in a clean manner, if stochasticity could be ignored.

\subsection{ Third Law }

The third law is a relationship law. It allows us to define the connection of the current level of description to its lower and higher ones. It is a reflection of the hierarchical structure of the whole dynamics. The essence of this law is to acknowledge the existence of two time scales: the macro time scale with which we observe the system dynamics at the focused level of description and the micro time scale with which fine structures and lower level dynamics come into play.

Specifically, it may be stated as follows: 
The Wright fitness function $\psi({\bf q},t)$ has the contribution from lower level in terms of time average on the time scale of current level, the contribution from the interaction among various components of the current level, and the contribution from higher level.
The stochastic drive $\xi({\bf q},t)$ is the remainder of all those contributions whose dynamics is fast on the time scale of current focus. Hence its average in time is zero. This stochastic contribution may be either unknown from a more fundamental level or unnecessary to be specified in details. Only its probability distribution is needed and is approximated by a Gaussian distribution in the present article.
The stochastic drive determines the ascendant matrix $A({\bf q},t)$, and the transverse matrix $T({\bf q},t)$ should be further determined by the dynamics of the system.

The lower level contribution to the Wright fitness function $\psi({\bf q}, t)$ and the stochastic drive $\xi({\bf q},t)$ may allow us to compute the intrinsic fitness landscape and the intrinsic source of evolution. However, historically the computation of this contribution tends to neglect the horizontal interaction among different components, which is usually nonlinear. 
On the other hand, the same and higher level contributions may suggest that a control mechanism, such as a feedback, may be from both of them in a large perspective. The combination of all three of them suggests that the evolution is nonlinear, asymmetric, mutually interactive, and stochastic, and may be controllable.
  
There is a degree of uncertainty and arbitrariness in the assignment of different levels of descriptions and the dichotomy of deterministic and stochastic terms in Eq.(1). This dilemma has been amply discussed in both physical \cite{vankampen} and biological \cite{turelli} sciences.
This has also been reflected in the mathematical theory of stochastic process.  Our way to solve this problem will be proposed in next section in connection to usual dynamics, which is different from both Ito and Stratonovich approaches and may be interpreted as an indication of the richness of the hierarchical structure.

\subsection{ Fundamental Theorem of Evolution }

In his classic treatise on genetical foundation of evolution \cite{fisher}, Fisher stated his fundamental theory of natural selection: {\it The rate of increase in fitness of any organism at any time is equal to its genetic variance in fitness at that time}.  This is truly a remarkable insight: It immediately connects the adaptation in fitness landscape to the stochastic drive. 

As implied in the first law, the ascendancy of the system is described by the ascendant matrix $A$, which in turn is completely determined by the stochastic drive according to the stochasticity-ascendancy relation, Eq.(2). The discussion followed Eq.(5) indicates that the ability of system to find a better fitness peak, not only the local fitness peak, or, to reach the global equilibrium, is guaranteed by the stochastic drive. This suggests that Eq.(2) is a statement on the unification of the two completely opposite tendencies: adaptation and randomization. 

Ever since Fisher's proposal of the fundamental theory of natural selection, misrepresentation and misunderstanding have been associated with this insightful statement, as discussed by Crow \cite{crow} and  Grafen \cite{grafen}. Though Fisher might have confined his discussion within genetic variations, the comparison of Fisher's statement to Eq.(2) points to a remarkable similarity: On the left hand side of Eq.(2) is the measure of the variation, and on the right hand side may be interpreted as the rate to fitness peak with a proper choice of unit.  Hence, we may also call Eq.(2) the fundamental theorem of evolution in a tribute to Fisher's great insight, or simply the Fisher's theorem. Its importance is evident: It is an integral part of the second law of evolution, the Darwinian dynamics. It enables the system to find the global maximum fitness peak. Nevertheless, it should be pointed out that the fundamental theorem of evolution in the present formulation is independent of the Wright fitness function.   

{\it Note:  In Fisher's original formulation \cite{fisher} he listed several exceptions to his fundamental theorem. According to the sharp analysis of S.J. Gould in his The Structure of Evolutionary Theory (2002) Fisher's exceptions completely negate the importance of Fisher's theorem! The present author believes that Fisher's misinterpretation of his theorem is the source of long confusion and debate in population genetics continuing to these days. See for example a most recent text book \cite{ewens}. In the present manuscript the precise mathematical formulation and clear ascendant matrix interpretation make all Fisher's exceptions unnecessary. Hence, the present formulation retains the fundamental nature of this theorem. 

The debate on Fisher's theorem is interesting even from physics point of view: There has been an enormous amount of confusion and debate on something similar in non-equilibrium thermodynamics. Perhaps the above formulation and the demonstration of the equivalence in the next section have been done in physics literature. Suggestions on relevant literature from readers would be highly appreciated. } 

\section{ Conventional Formulation }

\subsection{ Standard Stochastic Differential Equation }

Now we make the connection between the dynamics described by Eq.(1) and (2) to the dynamical equations typically encountered in evolution.
We start with the standard stochastic differential equation: 
\begin{equation}
  \dot{\bf q}_{t} = {\bf f}({\bf q}_t, t) + 
                    {\bf \zeta}( {\bf q}_t , t) \; . 
\end{equation}    
Here ${\bf f}({\bf q}, t)$ is the deterministic nonlinear drive of the system, which includes effects from both other components and itself, and the stochastic drive is ${\bf \zeta}( {\bf q}, t)$, which differs from that in Eq.(1) but is governed by the same dynamics. 
For simplicity we will assume that ${\bf f}$ is a smooth function whenever needed. Without loss of generality the stochastic drive in Eq.(8) is assumed to be Gaussian and white with the variance, 
\begin{equation}
 \langle {\bf \zeta}( {\bf q}_t , t) 
         {\bf \zeta}^{\tau}({\bf q}_{t'}, t') \rangle 
     = 2 D( {\bf q}_t, t) \; \epsilon \; \delta (t-t') ,
\end{equation}
and zero mean, $\langle {\bf \zeta}({\bf q}_t, t)\rangle = 0$. Eq.(9) is consistent with Eq.(2). Again here $\langle ... \rangle$ indicates the average with respect to the dynamics of the stochastic drive. According to the biology convention the semi-positive definite symmetric matrix $D=\{D_{ij}\}$ with $i,j=1,2, ..., n$ is the diffusion matrix.
Both the divergence and the skew matrix of the nonlinear drive ${\bf f}$ are in general non-zero:
\begin{equation}
 \nabla \cdot {\bf f} \neq 0, \; 
 \nabla \times {\bf f} \neq 0 \;.
\end{equation}
Here the divergence is explicitly $\nabla \cdot {\bf f} =  \sum_{j=1}^{n} \partial f_j /\partial q_j = tr(F)$, and the skew matrix $\nabla \times {\bf f}$ is twice the anti-symmetric part of the selection matrix $S$: $ ( \nabla \times {\bf f} )_{ij} = S_{ji} - S_{ij}$ with $ S_{ij} = \partial f_i /\partial q_j \ , \; i,j  =1,2, ..., n$.
The non-zero of the divergence leads to that the state space volume is not conserved: Ascendancy is implied. 
The non-zero of the skew matrix, or the asymmetry of the selection matrix $S$, implies the existence of the transverse matrix $T$. 

Now, we give an explicit construction which demonstrates the existence and uniqueness connection between Eqs. (1,2) and Eqs. (8,9). Assuming that both Eq.(1) and (8) describe the same dynamics. The speed $\dot{\bf q}_t $ is then the same in both equations. The connection from Eq.(1) to (8) is straightforward: Multiplying both sides of Eq.(1) by $ [A({\bf q}_t, t)  + T({\bf q}_t, t) ]^{-1} $ leads to Eq.(8). Converting Eq.(8) into (1) is mathematically more involved.
 
Using Eq.(8) to eliminate the speed in Eq.(1), and noticing that the dynamics of noise and the state variable behave independently, we have 
\begin{equation}
  [A({\bf q}, t) + T({\bf q}, t)] {\bf f}({\bf q}, t) 
   =  \nabla \psi({\bf q}, t) \; ,
\end{equation}
and 
\begin{equation}
   [A({\bf q}, t) + T({\bf q}, t)] {\bf \zeta}({\bf q}, t) 
       = \xi({\bf q}, t) \; . 
\end{equation}
Here we have dropped the subscript $t$ for the state variable, because time $t$ is now a parameter.
Those two equations suggest a rotation in state space.

Multiplying Eq.(12) by its own transpose of each side and carrying out the average over stochastic drive, we have
\begin{equation}
 [A({\bf q}, t) + T({\bf q}, t) ] D({\bf q}, t) 
 [A({\bf q}, t) - T({\bf q}, t) ] = A({\bf q}, t) \; .
\end{equation}
In obtaining Eq.(13) we have also used Eq.(2) and (9).
Eq.(13) suggests a duality between the standard stochastic differential equations and Eq.(1): A large ascendant matrix implies a small diffusion matrix. It is a generalization of the Einstein relation in the case of zero transverse matrix \cite{einstein}.

Next we define an auxiliary matrix function 
\begin{equation}
   G({\bf q}, t) = [A({\bf q}, t) + T({\bf q}, t) ]^{-1} \; .
\end{equation}
Here the inversion `${-1}$' is with respect to the matrix. Using the 
property of the Wright fitness function $\psi$: 
$\nabla \times \nabla \psi = 0$ [$(\nabla \times \nabla \psi)_{ij} = (\nabla_i \nabla_j -\nabla_i \nabla_j)\psi $ ], 
Eq.(11) leads to
\begin{equation}
  \nabla \times [ G^{-1} {\bf f}({\bf q}) ] = 0 \; ,
\end{equation}
which gives $n(n-1)/2$ conditions.
The generalized Einstein relation, Eq.(13), leads to the following 
equation
\begin{equation}
   G + G^{\tau} = 2 D \; ,
\end{equation}
which readily determines the symmetric part of the auxiliary matrix 
$G$, another $n(n+1)/$ conditions. The auxiliary function may be formally solved as an iteration in gradient expansion:
\begin{equation}
  G = D + Q \; , 
\end{equation}
with $Q = \lim_{j \rightarrow \infty} \Delta G_j $, 
$\Delta G_j = \sum_{l=1}^{\infty} (-1)^l [ (S^{\tau})^l 
 \tilde{D}_j S^{-l} + (S^{\tau})^{-l} \tilde{D}_j S^l ] $,
$\tilde{D}_0 = DS - S^{\tau}D$, 
$\tilde{D}_{j \geq 1} = ( D + \Delta G_{j-1} ) 
     \left\{ [\nabla \times (D^{-1} + \Delta G_{j-1}^{-1} ) ] 
     {\bf f} \right\} ( D - \Delta G_{j-1} )$. 
At each step of solving for $\Delta G_j$ only linear algebraic equation is involved. One can verify that the matrix $Q$ is anti-symmetric. For a simple case a formal solution of such algebraic equation was given in \cite{ao2002}, and an explicitly procedure was found for generic cases in \cite{kat}. Eq.(17) is a result of local approximation: If the selction matrix $S$, the diffusion matrix $D$ are constant in space, the exact solution only contains the lowest order contribution in gradient expansion: $Q= \Delta G_j = \Delta G_0 $. We regard Eq.(17) as the biological solution to Eq.(15) and (16), because it preserves all the fixed points of deterministic drive ${\bf f}$. 
The connection from Eq.(8) to (1) is therefore uniquely determined: 
\begin{equation}
  \left\{ \begin{array}{lll}
  \psi({\bf q}, t) & = & \int_C d{\bf q}' \cdot 
          [ G^{-1}({\bf q}') {\bf f}({\bf q}') ] \\
  A({\bf q}, t) & = & [G^{-1}({\bf q})+(G^{\tau} )^{-1}({\bf q})]/2 \\
  T({\bf q}, t) & = & [G^{-1}({\bf q})-(G^{\tau} )^{-1}({\bf q})]/2
  \end{array} 
     \right. \; . 
\end{equation} 
Here the sufficient condition $\det(A+T)\neq 0$ is used, and the end and initial points of the integration contour $C$ are $\bf q$ and ${\bf q}_0$ respectively.

We point out that in the absence of stochastic drive, i.e., $\epsilon = 0$ in Eq.(2) and (9), above connection remains unchanged.

\subsection{ Fokker-Planck Equation }

In many experimental studies in biology, a question is often asked on the distribution of the state variable as a function of time instead of focusing on the individual trajectory of the system. This implies that either there is an ensemble of identical systems or repetitive experiments are carried out. To describe this situation, we need a dynamical equation for the distribution function in the phase space. This goal can be accomplished by the so-called Fokker-Planck equation, or the difussion equation, or the Kolmogorov equation \cite{li,burger,vankampen}. 

In this subsection, another procedure to find the equation for distribution function is presented. It is natural from a theoretical physics point. This procedure will establish that the Wright fitness function $\psi$ in Eq.(1) indeed plays the role of potential energy in the manner envisioned by Wright, and the steady state distribution will be indeed given by Eq.(3). Our starting point will be the second law, Eq.(1), not the standard stochastic differential equation, Eq.(8), from which most previous derivations started.

The existence of both the deterministic and the stochastic drives in Eq.(1) suggests that there are two well separated time scales in the system: the microscopic or fine time scale to describe the stochastic drive and the macroscopic or course time scale to describe the system motion. The former time scale is much smaller than the latter. This separation of time scales further suggests that the macroscopic motion of the system has an "inertial": it cannot response instantaneously to the microscopic motion. To capture this feature, we introduce a small constant inertial "mass" $m$ and a kinetic momentum vector ${\bf p}$ for the system. Our state space is then enlarged: It is now a $2n$-dimensional space. The dynamical equation for the system takes the form:
\begin{equation}
   \dot{\bf q}_t = {\bf p}_t /m \; , 
\end{equation}
which defines the kinetic momentum, and 
\begin{equation}
    \dot{\bf p}_t = - [A({\bf q}_t, t) + T({\bf q}_t, t) ] {\bf p}_t /m 
                         + \nabla \psi({\bf q}_t, t) 
                         + {\bf \xi}({\bf q}_t, t) \; ,
\end{equation} 
which is the extension of Eq.(1). We note that there is no dependent of ascendant matrix $A$ and the stochastic drive on the kinetic momentum ${\bf p}$. The Fokker-Planck equation in this enlarged state space can be immediately obtained \cite{vankampen}:
\begin{equation}
 \left\{ \partial_t + \frac{\bf p}{m} \cdot \nabla_{\bf q} 
           + \overline{\bf f} \cdot \nabla_{\bf p} 
  - \nabla_{\bf p}^{\tau} A \left[\frac{\bf p}{m} + \nabla_{\bf p} 
\right]
\right\} 
   \rho({\bf q}, {\bf p}, t) = 0 \; . 
\end{equation}
Here $\overline{\bf f} = {\bf p}^{\tau} T /m + \nabla_{\bf q} \psi $, and $t$, 
${\bf q}$, and ${\bf p}$ are independent variables. 
The subscripts in the $\partial$ and $\nabla$ indicate the differentiation with respect to indicated variable only. 
The stationary distribution can be found, when the Wright fitness function is time-independent and bounded above, as \cite{vankampen} 
\begin{equation}
 \rho({\bf q},{\bf p},t=\infty) = \frac{1}{\cal Z} \exp\left\{-\frac{ 
   {{\bf p}^2 }/{2m} - \psi({\bf q}) }{\epsilon } \right\} \; ,
\end{equation}
with ${\cal Z} = \int \prod_{i=1}^{n} d{ q_i} \prod_{i=1}^{n} d{ p_i} 
  \exp\{ - [ {\bf p}^2 /2m - \psi({\bf q}) ]/\epsilon \}$ the partition function in the extended state space. 
There is an explicit separation of state variable and its kinetic momentum in Eq.(22). The elimination of the momentum in the small mass limit will not affect this distribution. Hence, Eq.(22) confirms that the expected Boltzmann-Gibbs distribution, Eq.(3) from the Eq.(1) and (2), is the right choice. 

We proceed to outline the procedure to find the Fokker-Planck equation corresponding to Eq.(1) and (2) without the kinetic momentum ${\bf p}$. We first illustrate how to recover Eq.(1) from Eq.(19) and (20). 
In the limit of $m \rightarrow 0$, the fast dynamics of kinetic momentum ${\bf p}_t$ can always follows the motion of slow dynamics of state variable ${\bf q}_t$. Hence we may set $\dot{\bf p}_t = 0 $ in Eq.(20) and replace the kinetic momentum using Eq.(19), which is then Eq.(1) after moving the speed to the left-side of equation. 
For the Fokker-Planck equation, the explicit separation of the kinetic momentum and state variable in the stationary distribution gives the guidance on the procedure: The resulting Fokker-Planck equation must be able to reproduce this feature. The Fokker-Planck equation is then found as
\begin{equation}
  \partial_t \rho({\bf q},t) 
   = \nabla^{\tau} [ - {\bf f}({\bf q}) - \Delta{\bf f}({\bf q}) 
                     + D({\bf q}) \nabla ] \rho({\bf q}, t) \; ,
\end{equation}
with $\Delta {\bf f}$ the solution of the equation 
$ \nabla \cdot \Delta {\bf f} + \Delta {\bf f}\cdot \nabla \psi 
  - \nabla \cdot [ G T G^{\tau} \nabla \psi ] = 0$.
%
%
If the probability current density is defined as ${\bf j}({\bf q},t) 
\equiv ( {\bf f} + \Delta{\bf f} - D \nabla ) \rho({\bf q},t)$, the 
Fokker-Planck equation is a statement of the probability continuity: 
\begin{equation}
  \partial_t \rho({\bf q},t) + \nabla \cdot {\bf j}({\bf q},t) = 0 \; .
\end{equation} 
The stationary state corresponds to the condition 
 $ \nabla \cdot {\bf j}({\bf q}, t=\infty) = 0 $.
One may verify that the stationary distribution $\rho({\bf q}, t=\infty)$ in Eq.(3) is indeed the time independent solution of the Fokker-Planck equation: The stationary probability current 
\begin{equation}
 {\bf j}({\bf q}, t=\infty) = (G T G^{\tau} + \Delta {\bf f}) \nabla \psi({\bf q}) \; \rho({\bf q},t=\infty) \; ,
\end{equation}
and $\nabla \cdot {\bf j}({\bf q},t=\infty) = 0 $.

The connection between the standard stochastic differential equation and Fokker-Planck equation has been under intensive study by biologists, physicists, chemists, mathematicians, and others over last 70 years \cite{li,burger,felsenstein,vankampen}. However, there exists an ambiguity for the generic nonlinear situation \cite{vankampen,turelli}. We attribute this ambiguity to the asymptotic nature of the connection in which a procedure must be explicitly defined: Different procedures will in general lead to different results. Biologically, it is a statement on how the dichotomy of deterministic and stochastic drives is done, a genuine indication of the hierarchical nature of the dynamics.
What has been demonstrated in this subsection is one way of carrying out this procedure.

\subsection{ Detailed Balance Condition }

There is an important class of evolution dynamics in which the anti-symmetric matrix $Q = 0$. Under this condition, the transverse matrix $T=0$, and  $\Delta {\bf f} = 0$. The Fokker-Planck equation becomes
\begin{equation}
  \partial_t \rho({\bf q},t) 
   = \nabla^{\tau} [ - {\bf f}({\bf q})) 
                     + D({\bf q}) \nabla ] \rho({\bf q}, t) \; ,
\end{equation}
and the stationary probability current is everywhere zero in state phase:
\begin{equation}
   {\bf j}({\bf q}, t=\infty) = 0 \; .
\end{equation}
In this situation one may find that
\begin{equation}
   \nabla \psi({\bf q})= D^{-1}({\bf q}) {\bf f}({\bf q}) \; ,
\end{equation}
and $A = D^{-1} $.
The Wright fitness function and the connection between Eq.(1) and the standard stochastic differential equation can be directly read out from equations. This is the well-known symmetric dynamics in biological \cite{li,burger,felsenstein} and physical \cite{vankampen} sciences. 
A condition to generate this kind of equilibrium state in biology was first noticed by Hardy \cite{hardy} and Weinberg \cite{weinberg}. This zero probability current condition is usually called the detailed balance condition.  

\section{ Examples }

In this section we discuss four examples. The first one is of predator-prey model like. In this model we illustrate how to approximately compute the Wright fitness function $\psi$, the ascendant matrix $A$ and the transverse matrix $T$, that is, how to make the connection between the conventional formulation and Eq.(1) and (2). Second and third examples are current models in evolutionary biology. Fourth example is the usual diffusion equation. We use them to demonstrate that they may be discussed within the present mathematical formulation. 

\subsection{ Predator-Prey Model }
   
The example whose dynamical equation is in the form of standard stochastic differential equation is the generic predator-prey process. Under the diffusion approximation, both the diffusion matrix $D$ and the deterministic drive can be obtained from the master equation. The diffusion approximation is valid when a large number of birth and death events occurs on the macroscopic time scale \cite{vankampen}.

We remark that in reality, both the intrinsic and the extrinsic noise coexist. They are equally important and measurable \cite{vankampen}. 
The stochasticity-ascendancy relation, or the fundamental theorem of evolution, treats both of them on the equal footing to determine the connection between Eq.(1) and (8).

We now give an explicit demonstration of how to obtain Eq.(1) from Eq.(8). Here $q_1$ and $q_2$ represent numbers of two species in a habitat. 
We assume the spatial distribution is uniform. The deterministic drive ${\bf f}$ consists of two positive terms, birth and death:
\begin{equation}
    f_i({\bf q}) = f_{ib}({\bf q}) - f_{id}({\bf q}) \; \; i=1,2 \; ,
\end{equation}
with the subscripts $b$ and $d$ stand for the birth and death respectively. Under the diffusion approximation, the stochastic drive is \cite{li,mcquarrie,vankampen}
\begin{equation}
 \zeta_i({\bf q},t) = \sqrt{ f_{ip}({\bf q}) } \zeta_{ip}(t) +
                      \sqrt{ f_{id}({\bf q}) } \zeta_{id}(t)
                         \; \; i=1,2 \; ,
\end{equation}
with $\zeta_{ip}(t), \zeta_{id}(t)$ are unity random variables and possible correlation among them. Therefore the diffusion matrix $D$ can be readily obtained, which is what needed below. We remark that the equation similar to predator-prey equation has been emerged in the study of the robustness of the gene regulatory network of phage $\lambda$ \cite{zhu}.

The construction of Eq.(1) from Eq.(8) will be given to the lowest order in the gradient expansion. The usefulness of this approximated construction can be illustrated for following two reasons. First, in many practical applications, lowest order approximation is already enough \cite{zhu}, because it is exact in the strictly linear case. Second, several salient features of the connection becomes apparent without undue mathematical complications. An important quantity is the selection matrix $S$. According to the definition following Eq.(13),
\begin{equation}
    S_{11} = \nabla_1 f_1 \, ,    S_{12} = \nabla_2 f_1 \, ,
    S_{21} = \nabla_1 f_2 \, ,    S_{22} = \nabla_2 f_2  \; .
\end{equation}
Eq.(10) will not change under the gradient approximation. In the lowest order gradient approximation, Eq.(9) becomes simple. We collect them here: 
\begin{equation}
  \left\{ \begin{array}{lll}
   G S^{\tau} - S G^{\tau} & = & 0  \\
   G          + G^{\tau}   & = & D 
  \end{array} 
     \right. \; . 
\end{equation} 
In two dimensions the matrix manipulation is particularly straightforward. We note that any $2\times 2$ matrix $M$ can uniquely decomposed in terms of Pauli matrices, $\sigma_i$ with $i=1,2,3$, and the identical matrix ${\bf 1}$:
\[
  M = M_1 \sigma_1 + M_2 \sigma_2 + M_3 \sigma_3 
      + tr(M)/2 \; {\bf 1} \; ,
\]
with $tr$ denotes the trace and $\sigma_1 = \left( \begin{array}{ll}
                 0 & 1  \\
                 1 & 0 
    \end{array} \right) $, 
    $\sigma_2 = \left( \begin{array}{ll}
                 0 & -i  \\
                 i & 0 
    \end{array} \right) $, 
    $\sigma_1 = \left( \begin{array}{ll}
                 1 & 0  \\
                 0 & -1 
    \end{array} \right) $, and here $i = \sqrt{-1}$.  
Using this relationship, the equation for antisymmetric part of the auxiliary matrix $G = D + Q$ from Eq.(32) is
\begin{equation}
  Q S^{\tau} + S Q =  (S D - D S^{\tau}) \; .
\end{equation} 
Using the matrix decomposition and the properties of Pauli matrices, 
we obtain
\begin{equation}
  Q = { (S D - D S^{\tau}) }/ { tr(S) } \; .
\end{equation} 
Note that for the $2\times2$ matrix $M$
\[
   \left( \begin{array}{ll}
                 M_{11} & M_{12} \\
                 M_{21} & M_{22}  
    \end{array} \right) ^{-1} = \frac{1}{ \det(M) }
    \left( \begin{array}{ll}
                 M_{22} & - M_{12} \\
               - M_{21} & M_{11}  
    \end{array} \right) 
\]
The ascendant matrix $A$ and the transverse matrix $T$ can be found according to Eq.(8):
\begin{equation}
  \left\{ \begin{array}{lll}
    \psi({\bf q}) & = & \int_C d{\bf q}' \cdot 
          [ G^{-1}({\bf q}') {\bf f}({\bf q}') ] \\
    A({\bf q}) & = &
         \left( \begin{array}{ll}
                 D_{22} & - D_{12} \\
               - D_{12} & D_{11}  
         \end{array} \right)  /\det(G)  \\
    T({\bf q}) & = & {-Q}/{\det(G)} 
  \end{array} 
     \right. \; . 
\end{equation} 
In two dimensions, $\det(G) = \det(D) + \det(Q)=D_{22}D_{11}-D_{12}^2 + Q_{12}^2 $ and is obviously non-negative.

To summarize, the computation of quantities in Eq.(1) from Eq.(8) is as follows: First, to establish Eq.(8) from the biological problem, continuous approximation should be used. If the problem is given in terms of master equation, the usual diffusion approximation will be employed, which gives both the diffusion matrix and the deterministic drive \cite{burger,vankampen}. After this is done, the procedure prescribed in section III.A is employed to find the ascendant matrix, the transverse matrix and the fitness matrix. According to the biological problem, an additional approximation may be used to reduce computation effort but with the desired accuracy, as done here as well as in Zhu {\it et al.} \cite{zhu}.

\subsection{ Bateson-Dobzhansky-Muller Model }

Based on the verbal descriptions of Bateson, Dobzhansky, and Muller,
Gavrilets has developed a mathematical description of adaptive landscape with many loci and traits to describe the speciation: the holey adaptive landscapes \cite{gavrilets}. 
It puts the emphasis on narrow fitness bands linking large fitness areas with the same fitness. The fitness peaks, expressed at rugged adaptive landscapes, are argued to be not important when number of traits and loci is large, contrast to the demonstration by Kisdi and Geritz \cite{kisdi}. 

This is a very attractive idea, and is consistent with present framework. With an appropriate coarse grain average, it may be possible to compute this holey adaptive landscape in terms of the Wright fitness function defined in the present article.
The speciation may be interpreted as the diffusion along the narrow fitness bands, a process may be much faster than that described by 
Eq.(5) for the peak hopping.
   
\subsection{ Symmetry-Breaking Model }

The symmetric evolution dynamics was explicitly discussed by Stewart for speciation \cite{stewart}:
\begin{equation}
  \dot{\bf q }_t = \nabla \phi({\bf q}_t ) \; .
\end{equation}
This is a deterministic equation. Stochastic model was also mentioned by Stewart \cite{stewart}. 

In the light of present discussion, the diffusion matrix $D$ is equivalent to the identity matrix ${\bf 1}$. This is a case satisfying the detailed balance condition. Hence the Wright fitness function can be directly obtained: $\psi({\bf q}_t ) = \phi({\bf q}_t )$, and the ascendant matrix $A= {\bf 1}$, and transverse matrix $T=0$.
The steady state distribution is then given by the Boltzmann-Gibbs like distribution, Eq.(3). All the statistical physics methodology can then be applied here. Naturally one may make use of the idea of symmetry-breaking as a way for self-organization needed for speciation. We refer readers to the beautiful discussion presented by Stewart \cite{stewart}.  

We remark that one may even make use of the self-consistent mean-field approximation, a powerful mathematical tool in statistical physics \cite{goldenfeld}, to search for the indication of symmetry-breaking.

\subsection{Diffusion Equations}

Expressing evolutionary dynamics in the form of diffusion equations has been common in population genetics \cite{burger,felsenstein}.
Particularly, the one-dimensional diffusion equation has been thoroughly studies. It belongs to the class satisfying the detailed balance condition. Hence, as discussed in subsection III.C, the transverse matrix is zero, and the ascendant matrix and the gradient of the Wright fitness function, or adaptive function, can be readily identified.
The results, such as the equilibrium population, are consistent with what known.  We remark that if there is an overall factor difference in terms of the diffusion matrix, one needs to examine which dichotomy of deterministic and stochastic drives is used:
Ito, Stratonovich, the present prescription, or something else.
A discussion of this feature can be found in the monograph of van Kampen \cite{vankampen}.

In case the detailed balance condition would not hold for higher dimensional diffusion equations, a way to proceed along the present formulation is discussed at the end of subsection IV. A.

\section{ Discussions }

Before further going to further discussion on the implication of the present mathematical formulation, it should be kept in mind that above three laws must be regarded as mathematically what the evolutionary dynamics might be. They are by not means the exact description. Having made this statement, we nevertheless remark that although proposed three laws for evolutionary dynamics are based on the continuous approximation in terms of time and population, it is possible that main features discussed in the present article may survive in discrete cases.

In the present attempt to unify approaches from biology and physical sciences, the possible existence of laws such as expressed by Eq.(1) and (2) should not be too surprised if one views them from two important principles which have been rigorously validated: 
1) Simple equations can generate extremely complicated patterns and phenomena \cite{may,murray,guckenheimer}; 
2) Each level of description has its own laws which cannot be derived in a naive reductionism manner \cite{anderson,goldenfeld,batterman}. 
The connections between levels are asymptotic and emerging phenomena frequently occur at higher levels.   

There are several quantitative advantages in present formulation of evolutionary theory. With the Wright fitness function defined as in Eq.(1), an independent way to calculate it is obtained, as one may follow what indicated in Section III. This adds more predictive power into the evolutionary dynamics. 

As expressed by Eq.(2) and discussed in section 2.4, Fisher's fundamental theorem of natural selection (Fisher's theorem) becomes transparent and indispensable. This may provide a much-needed step to better understand Fisher's great insight. 

Combining both Fisher's and Wright's insights, the Wright fitness function and Fisher's theorem provide a quantitative measure to discuss robustness, stability, and the speciation. Eq.(5) is such an example.

It is interesting to point out the remarkable similarity between the adaptive landscape of Wright \cite{wright} and the developmental landscape of Waddington \cite{waddington}. The present mathematical formulation can deal with both cases. An example on the gene regulatory network in phage $\lambda$ has already been studied by Zhu {\it et al.} \cite{zhu}. This suggests a unification between genetics and developmental biology.

However, one may ask why to use Eq.(1) and (2) instead of more conventional Eq.(8) and (9): After all their equivalence has been demonstrated above. 
Here we offer three reasons to favor Eq.(1) and (2):
 
i) Quantities presented in Eq.(1) can be directly related to experimental observation. For example, Eq.(3) gives a direct connection between the Wright fitness function and the population density in steady state. By observing the dynamical behaviors, information on the ascendant and transverse matrices can be obtained. 
Also, Eq.(5) can relate stability to the Wright fitness function. This direct contact with experimental data is an indication of the autonomy of the focused level of description.

ii) Eq.(8) and (9) lack the visualizing ability for the global dynamics behavior. For example, in a nonlinear dynamics with multiple local maxima, it is not clear from Eq.(8) and (9) which maximum is the largest one, and how easy it might be to mover from one maximum to another.
One could find this answer by a direct real time calculation. 
But this is usually computationally demanding, if not impossible.

iii) Eq.(1) and (2) give an alternative modeling of evolutionary dynamics, which can be advantageous in certain situations. For example, the direct use of fitness function in Stewart's modeling \cite{stewart} makes the symmetry-breaking idea very transparent from statistical physics' point of view.  

Finally, we point out an interesting mutual reduction loop between biology and physics. In the discussion of the first law we have remarked that one may regard the Newtonian dynamics as a special case of the first law. This implies that the Newtonian conservative dynamics is a special case of the present second law, hence the Darwinian dynamics. Here the opposite statement also exists: Under an appropriate condition equations in the form of Eq.(1) and (2) can be derived from the Newtonian dynamics \cite{leggett,at,az}, therefore the Darwinian dynamics may also be regarded as a special case of the Newtonian dynamics.  

\section{ Conclusions }

Based on continuous approximation we have postulated three laws to mathematically describe the evolutionary dynamics. The most fundamental equation, the second law, has been expressed in a unique form of stochastic differential equation. Four dynamical elements have used in our formulation: the ascendant matrix, the transverse matrix, the Wright fitness function, and the stochastic drive. 
We have demonstrated that present laws are consistent with previous approaches in biology, but appear more suitable to discuss stability and other phenomena quantitatively. 
Various important results, such as Fisher's fundamental theorem of natural selection and Wright's adaptive landscape, as well as the developmental landscape, are apparently unified in the present formulation.
It appears that the present quantitative formulation has captured the main aspects of the Darwinian dynamics. 

{\ }

\noindent{\bf Acknowledgement:} The critical comments of J. Felsenstein and D. Waxman are highly appreciated. I also appreciate very helpful discussions with C.T. Bergstrom, L. Yin, X.-M. Zhu, Z. Chen, C. Kwon, H. Qian, and D.J. Thouless for various aspects on the underlying mathematics.  
This work was supported in part by a USA NIH grant under HG002894-01.
 
{\ }


\begin{thebibliography}{99}

\bibitem{grene}
 Grene, M. (1983) (ed) Dimensions of Darwinism: themes and
  counterthemes in twentieth century evolutionary theory. 
  Cambridge University Press, Cambridge.
\bibitem{kauffman}
 Kauffman, S.A. (1993) The origins of order, Oxford University Press,
   Oxford.  
\bibitem{epistasis}
 Wolf, J.B., Brodie III, E.D., and Wade, M.J. (2000) (ed) Epistasis and 
  the evolutionary process. Oxford University Press, Oxford. 
\bibitem{coyne}
 Coyne, J.A., Barton, N.H. and Turelli, M. (2000) Is Wright's shifting
   balance process important in evolution? Evolution. 54: 306-317.
\bibitem{goodnight}
 Goodnight, C.J. and Wade, M.J. (2000) The ongoing synthesis: a reply  
   to Coyne, Barton, and Turelli. Evolution. 54: 317-324.  
\bibitem{peck}
 Peck, S.L., Ellner, S.P. and Gould, F. (200) Varying migration and
   deme size and the feasibility of the shifting balance. 
    Evolution. 54: 324-327.   
\bibitem{pigliucci}
 Pigliucci, M. (2003) Species as family resemblance concepts: 
   the (dis-)solution of the species problem?   Bioassay, 25: 596-602.
\bibitem{crow}
 Crow, J.F. (2002) Here's to Fisher, additive genetic variance, and the 
  fundamental theorem of natural selection. Evolution, 56: 1313-1316.  
\bibitem{grafen}
 Grafen, A. (2003) Fisher the evolutionary biologist, 
  The Statistician, 52: 319-329.
\bibitem{stewart}
 Stewart, I. (2003) Self-organization in evolution: a mathematical  
   perspective. Phil. Trans. R. Soc. Lond. A361: 1101-1123.
\bibitem{gavrilets}
 Gavrilets, S. (2003) Models of species: what have we learnt in 
   40 years? Evolution, 57: 2197-2215.
\bibitem{kisdi}
 Kisdi, E. and Geritz, S.A.H. (1999) Adaptive dynamics in allele space: 
  evolution of genetic polymorphism by small mutations in a
  heterogeneous environment. 53 993-1008.
\bibitem{es1998}
 van der Vijver, G., Salthe, S.N. and Delpos, M. (1998) (ed)
  Evolutionary systems: biological and epistemological perspectives on 
  selection and self-organization. Kluwer, Dordrecht.
\bibitem{li}
  Li, W.H. (1977) (ed) Stochastic models in population genetics.
   Dowden, Hutchison, and Ross, Inc., Stroudsburg.
\bibitem{burger}
  Burger, R. (2000) The mathematical theory of selection, 
    recombination, and mutation. John Wiley, New York.
\bibitem{ewens}
  Ewens, W.J. (2004) Mathematical population genetics: I. Theoretical
    introduction. 2nd edition. Springer, Berlin.   
\bibitem{felsenstein} 
  Felsenstein, J. (2003) Theoretical evolutionary genetics. 
    Online book. 
  (http://evolution.genetics.washington.edu/pgbook/pgbook.html)
\bibitem{maynardsmith}
 Maynard Smith, J. (1982) Evolution and the theory of games. Cambridge 
  University Press, Cambridge. 
\bibitem{wright}
 Wright, S. (1932) The roles of mutation, inbreeding, crossbreeding 
  and selection in evolution. Proceedings of the Sixth International
  Congress of Genetics, 1: 356-366. 
\bibitem{kramers}
 Kramers, H.A. (1940). Brownian motion in a field of force and the 
  diffusion model of chemical reactions. Physica 7: 284-304.
\bibitem{barton}
 Barton, N.H. and Rouhani, S.R. (1987) The frequency of shifts between
   alternative equilibria. J. Theor. Biol. 125: 397-418.  
\bibitem{vankampen}
 Kampen, van N.G. (1992). Stochastic processes in physics and
  Chemistry. Elsevier, Amsterdam. 
\bibitem{darwin1858}
 Darwin, C. and Wallace, A. (1858) On the tendency of species to form
  varieties; and on the perpetuation of varieties and species by  
  natural means of selection. Journal of the Proceedings of the Linnean
  Society of London, Zoology 3: 45-62.   
\bibitem{darwin1958}
 Darwin, C. (1958) On the origin of species by means of natural 
  selection, or the preservation of favoured races in the struggle for
  life. Penguin, New York. 
\bibitem{founders}
 Sarkar, S. (1992) (ed) The founders of evolutionary genetics: a
   centenary reappraisal. Kluwer Academic Publishers, Dordrecht.
\bibitem{jacob}
 Jacob, F. (1982) The possible and the actual. University of
     Washington Press, Seattle.
\bibitem{monod}
 Monod, J. (1971) Chance and necessity: an essay on the natural 
     philosophy of modern biology. Knopf, New York.
\bibitem{kimura}
 Kimura, M (1983) The neutral theory of molecular evolution. Cambridge
   University Press, Cambridge.
\bibitem{guckenheimer}
 Guckenheimer, J. and Holmes, P. (1997) Nonlinear oscillations,
  Dynamical systems, and bifurcations of vector fields.
  Springer-Verlag, Berlin.
\bibitem{murray}
 Murray, J.D. (2002) Mathematical biology. v. 1. Springer, New York.
\bibitem{may}
 May, R.M. (1981) (ed) Theoretical ecology: principles and
   applications. Second edition. Blackwell Scientific, Oxford.
\bibitem{goldstein}
 Goldstein, H. (1980) Classical mechanics. Second edition. 
  Addison-Wesley, Reading.
\bibitem{dobzhansky}
 Dobzhansky, T.G. (1971) Genetics of the evolutionary process. Columbia
  University Press, New York.
\bibitem{mayr}
 Mayr, E. (1982) The growth of biological thought: diversity,
  evolution, and inheritance.  Harvard University Press, Cambridge. 
\bibitem{turelli}
 Turelli, M. (1977) Random environments and stochastic calculus. 
  Theor. Popul. Biol. 12: 140-178. 
\bibitem{fisher}
 Fisher, R.A. (1930) The genetical theory of natural selection,
  Clarendon, Oxford.
\bibitem{einstein} 
 Einstein, A. (1905) Ann. Physik 17: 549-56.
\bibitem{ao2002}
 Ao, P. (2002) Stochastic force defined evolution in dynamical 
  systems. (http://it.arXiv.org/find/physics/1/Ao/0/1/0/past/3/0)
\bibitem{kat}
 Kwon, C., Ao, P. and Thouless, D. J. (2003). Structure of stochastic
  dynamics near fixed points (submitted to PNAS, available upon 
  request).
%
\bibitem{hardy}
 Hardy, G.H. (1908) Mendelian proportions in a mixed population.
  Science 28: 49-50.
\bibitem{weinberg}
 Weinberg, W. (1908) Uber den nachweis der vererbung beim menschen. 
  jahresch. Ver. Vaterl. Naturkel. Wuerttemb Stuttgart. 64: 368-382.     
\bibitem{mcquarrie}
 McQuarrie, D.A. (1967) J. Appl. Prob. 4: 413.
\bibitem{zhu}
 Zhu, X.-M., Yin, L., Hood, L. and Ao P. (2003) 
  Calculating biological behaviors of epigenetic states in phage
  $\lambda$ life cycle. Functional and Integrative Genomics 
  (DOI: 10.1007/s10142-003-0095-5, in press).
\bibitem{goldenfeld}
 Goldenfeld, N. (1992) Lectures on phase transitions and the
   renormalization group. Addison-Wesley, Reading.              
\bibitem{anderson}
  Anderson, P.W. (1972) More is different. Science 177: 393-396. 
\bibitem{batterman}
 Batterman, R.W. (2001) The devil in the details: asymptotic reasoning
   in explanation, reduction, and emergence. 
   Oxford University Press, Oxford.
\bibitem{waddington}
 Waddington, C.H. (1940) Organisers and genes. Cambridge University
    Press, Cambridge.
\bibitem{leggett}
 Leggett, A.J. (1992) Quantum tunnelling of a macroscopic variable.
  pp1-36. In Quantum tunnelling in condensed media. edited by Yu.
  Kagan, A.J. Leggett, North-Holland, Amsterdam. 
\bibitem{at}
 Ao, P. and Thouless, D.J. (1993) Berry phase and the Magnus force for
   a vortex line in a superconductor. Phys. Rev. Lett. 70: 2158-2161.  
\bibitem{az}
 Ao, P. and Zhu, X.-M. (1999) Microscopic theory of vortex dynamics in
   homogeneous superconductors. Phys. Rev. B60: 6850-6877.

\end{thebibliography}
\end{document}